\documentclass[conference]{IEEEtran}
\usepackage[left=0.64in,right=0.64in,top=0.75in,bottom=1.2in]{geometry}
\usepackage{algorithm}
\usepackage[noend]{algpseudocode}

\usepackage{type1cm}
\usepackage{lettrine}
\usepackage{amsthm}
\usepackage{amsmath}
\usepackage{amsfonts} 
\usepackage{bbm}
\usepackage[pdftex]{graphicx}
\usepackage{amsmath}
\usepackage[table,xcdraw]{xcolor}
\usepackage{colortbl}
\usepackage{multirow}
\usepackage{mathrsfs}
\usepackage{type1cm}
\usepackage{lettrine}
\usepackage[numbers, sort&compress]{natbib}
\usepackage{setspace}
\usepackage[normalem]{ulem}

\usepackage{hhline}
\usepackage{color, colortbl}
\usepackage{subcaption}

\usepackage{tabularx}
\newcolumntype{g}{>{\columncolor{lightgray!20}}c}

\usepackage{mathtools, stmaryrd}
\usepackage{xparse} \DeclarePairedDelimiterX{\Iintv}[1]{[}{]}{\iintvargs{#1}}
\NewDocumentCommand{\iintvargs}{>{\SplitArgument{1}{,}}m}
{\iintvargsaux#1} %
\NewDocumentCommand{\iintvargsaux}{mm} {#1\mkern1.5mu..\mkern1.5mu#2}

\newcounter{function}
\makeatletter

\newcommand{\vertices}{\tilde{\boldmath{U}}}
\newcommand{\vertex}{\tilde{u}}

\newcommand{\edges}{{\boldmath{E}}}
\newcommand{\edge}{{e}}

\newcommand{\weights}{{\boldmath{W}}}
\newcommand{\weight}{{w}_{{e}}}

\newcounter{constraint}
\newenvironment{constraint}[1][]
{\refstepcounter{constraint}
\begin{equation}\tag{C\theconstraint}}
{\end{equation}}

\begin{document}

\newtheorem{theorem}{Proposition}

\title{Optimizing Collaborative UAV Networks for Data Efficiency in IoT Ecosystems}


\author{%
\IEEEauthorblockN{Priyavrat Dev Sharma\IEEEauthorrefmark{1},
Ibrahim Sorkhoh\IEEEauthorrefmark{2},
Muthucumaru Maheswaran\IEEEauthorrefmark{3}}
\IEEEauthorblockA{\IEEEauthorrefmark{1}Electrical and Computer Engineering, McGill University, Canada}
\IEEEauthorblockA{\IEEEauthorrefmark{2}GEAR, Gulf University for Science and Technology, Kuwait}
\IEEEauthorblockA{\IEEEauthorrefmark{3}School of Computer Science, McGill University, Canada}
}
\maketitle

\begin{abstract}
Advances in the Internet of Things are revolutionizing data acquisition, enhancing artificial intelligence and quality of service. Unmanned Aerial Vehicles (UAVs) provide an efficient data-gathering solution across varied environments. This paper addresses challenges in integrating UAVs for large-scale data operations, including mobility, multi-hop paths, and optimized multi-source information transfer. We propose a collaborative UAV framework that enables efficient data sharing with minimal communication overhead, featuring adaptive power control and dynamic resource allocation. Formulated as an NP-hard Integer Linear Program, our approach uses heuristic algorithms to optimize routing through UAV hubs. Simulations show promise in terms of computation time (99\% speedup) and outcome (down to 14\% deviation from the optimal).

\end{abstract}
\section{Introduction}


The growing deployment of Internet of Things (IoT) devices has led to an explosion in collected data across various purposes. This vast amount of data can be leveraged for different applications through artificial intelligence (AI) tools and methods. Unmanned Aerial Vehicles (UAVs) have emerged as a cost-effective technology offering innovative solutions to data-gathering challenges across diverse IoT environments \citep{wei2022uav}.

UAVs possess powerful capabilities that make them well suited for collecting data in various scenarios, from urban landscapes to remote, hard-to-reach areas. Their high-speed mobility, deployment with various communication technologies (e.g., WiFi, 5G Cellular), and data acquisition capabilities enable them to gather data rapidly when working collaboratively. This collaborative approach allows for increased coverage, real-time data sharing, reduced communication load on base stations, enhanced collaborative capabilities, and improved resilience in case of communication loss with base stations.

Collaborative UAV networks can enhance various applications by gathering and sharing information. For environmental monitoring, UAVs gather data on urban air quality, temperature, and pollution levels, creating comprehensive maps with real-time updates for anomaly detection and hazard alerts. In disaster scenarios, thermal camera-equipped UAVs efficiently locate survivors and ensure fast and thorough area coverage. UAVs also transform agriculture by employing multispectral camera-equipped UAVs to survey fields to facilitate collaborative data exchange on crop health, pest issues, and irrigation requirements. The collaborative nature of these UAV networks significantly improves data accuracy and coverage in these use cases. By sharing information in real-time, UAVs can cross-validate their findings, fill gaps in individual data collection, and provide a more comprehensive and up-to-date picture of the situation.

Integrating UAVs into IoT and data gathering systems offers opportunities and challenges. Although UAVs can efficiently cover vast areas and access rugged terrain, they also bring new considerations such as limited power, payload capacity, and communication range. To fully realize this technology's potential in large-scale data-gathering operations,  it is crucial to develop strategies that optimize UAV-based data collection while minimizing communication overhead. Several challenges should be tackled to realize such a solution. 1) Although the high mobility of UAVs gives them advantages compared to other solutions, it represents a challenge when developing a communication plan. 2) Establishing multi-hub paths via different time steps requires a delicate system model to handle data transfer effectively. 3) Transferring information from multiple sources to multiple destinations is a challenging task, not to mention that there are multiple pieces of information to disseminate, utilizing the same limited number of devices and wireless communication resources. 

In this work, we present a data-gathering scheme using multiple UAVs that communicate to share the information collected from sources and deliver it to destinations while minimizing communication overhead. Our system incorporates multiple power levels, each with its corresponding communication range, and dynamically allocates wireless resources for data transmission. Our system model considers several UAVs mobilizing a confined space to monitor area status and share the collected information (see Section \ref{sec:systemModel}). We proved the problem is NP-hard through reduction and modeled it as an Integer Linear Program (ILP) (see Section \ref{sec:ILP}). We propose several heuristics that route the required information from its UAV source to its UAV destinations via multiple hub UAVs (see Section \ref{sec:heuristics}), demonstrating the application of collaborative UAV information collection and sharing in various fields. We have conducted several experiments to illustrate the efficiency of our proposed solution (see Section \ref{sec:perEval}). 


\section{Related Work} \label{sec:RelWork}

Energy efficiency is a critical concern in UAV-based systems due to the limited power resources of these aerial vehicles. 
In \citep{yang2019energy}, they investigated the problem of sum power minimization in a mobile edge computing network with multiple UAVs, jointly optimizing user association, power control, allocation of computation capacity, and location planning. Authors of \citep{kim2020energy} proposed an energy-efficient, cooperative multi-hop relay scheme for UAVs, formulating an optimal multi-hop transmission scheduling problem to minimize power consumption while meeting time constraints to reach the base station. The work in \citep{zhan2017energy} considered a general fading channel model for sensor node-UAV links, jointly optimizing the sensor nodes' wakeup schedule and UAV trajectory to minimize maximum energy consumption while ensuring reliable data collection. Efficient data collection is crucial the utility of UAV-based systems in IoT and wireless sensor networks. In \citep{meng2022multi}, they presented the optimization of mission completion time is mainly concerned with the formulation of the task allocation problem along with the optimization of transmit power. Also, the work in \citep{chen2017caching} proposed a novel algorithm based on the machine learning framework of conception-based echo state networks to report event packets in an energy-efficient manner while maximizing users' quality of experience and minimizing UAV transmit power.

Optimizing UAV deployment and network configuration is essential to improve overall system performance. The work in \citep{wang2023cooperative} minimized the information size in multi-UAV cooperative data collection by optimizing UAV trajectories and sensor transmission scheduling. In \citep{chen2022joint}, they combined and optimized the trajectories of the UAVs, total throughput, and energy efficiency by associating each user to a certain UAV. In this work, it was assumed that UAVs do not collaborate. In another study, in \citep{fu2024collaborative}, they developed a method to extend the lifetime of the maritime sensor network(MSN) by optimizing the hover points of UAVs and data collection nodes of subnets and they have shown that the data collection scheme can achieve low data transmission latency and significantly extend the lifetime of the MSN. In \citep{10313259}, authors presented a comprehensive model to combine UAV path and hovering time to enable uninterrupted and efficient data exchange among UAVs. 


Efficient path planning and trajectory optimization are crucial for maximizing the effectiveness of UAV-based data-gathering systems.  A trajectory optimization algorithm for multiple UAVs in data gathering scenarios, considering both energy consumption and data collection efficiency, was proposed in \citep{cai2018trajectory}. The work in \citep{liu2019trajectory} developed a joint trajectory and resource allocation optimization framework for UAV-enabled mobile edge computing networks, with the aim of minimizing the total energy consumption of the system. The concept of UAV swarms has gained attention for its potential to enhance data-gathering capabilities. Authors of \citep{campion2018uav} explored the use of UAV swarms for distributed data gathering in large-scale environments, proposing a decentralized coordination algorithm based on artificial potential fields. In \citep{yanmaz2018drone}, they investigated the challenges and opportunities of using drone swarms for various applications, including data gathering and communication in disaster scenarios. The integration of UAV-based systems with edge computing paradigms has emerged as a promising research direction. The work in \citep{zhou2020uav} proposed a UAV-enabled mobile edge computing system that jointly optimizes task offloading, resource allocation, and UAV trajectory to minimize the overall energy consumption and task completion time. Another work in \citep{hu2020optimization} developed a framework for UAV-assisted edge computing in Internet of Things (IoT) networks, addressing challenges such as task off-loading, resource allocation, and UAV positioning.

Our work addresses the unique challenges of collaborative UAV networks in data-gathering operations. We focus on mobility-induced communication planning, the establishment of multiple hop paths, and efficient multisource to multidestination information transfer (in contrast to \cite{kim2020energy}). Our approach integrates adaptive power modulation and dynamic resource allocation, which is crucial for optimizing UAV operations with limited power. We formulate the problem, prove its NP-hardness, and develop novel heuristic algorithms for routing via multiple UAV hubs. In this way, we combine theoretical analysis with practical solutions.

\vspace{-1mm}
\section{System Model} \label{sec:systemModel}
\begin{figure}[!htb]
    \vspace{-5mm}
    \centering
    \includegraphics[width =\columnwidth]{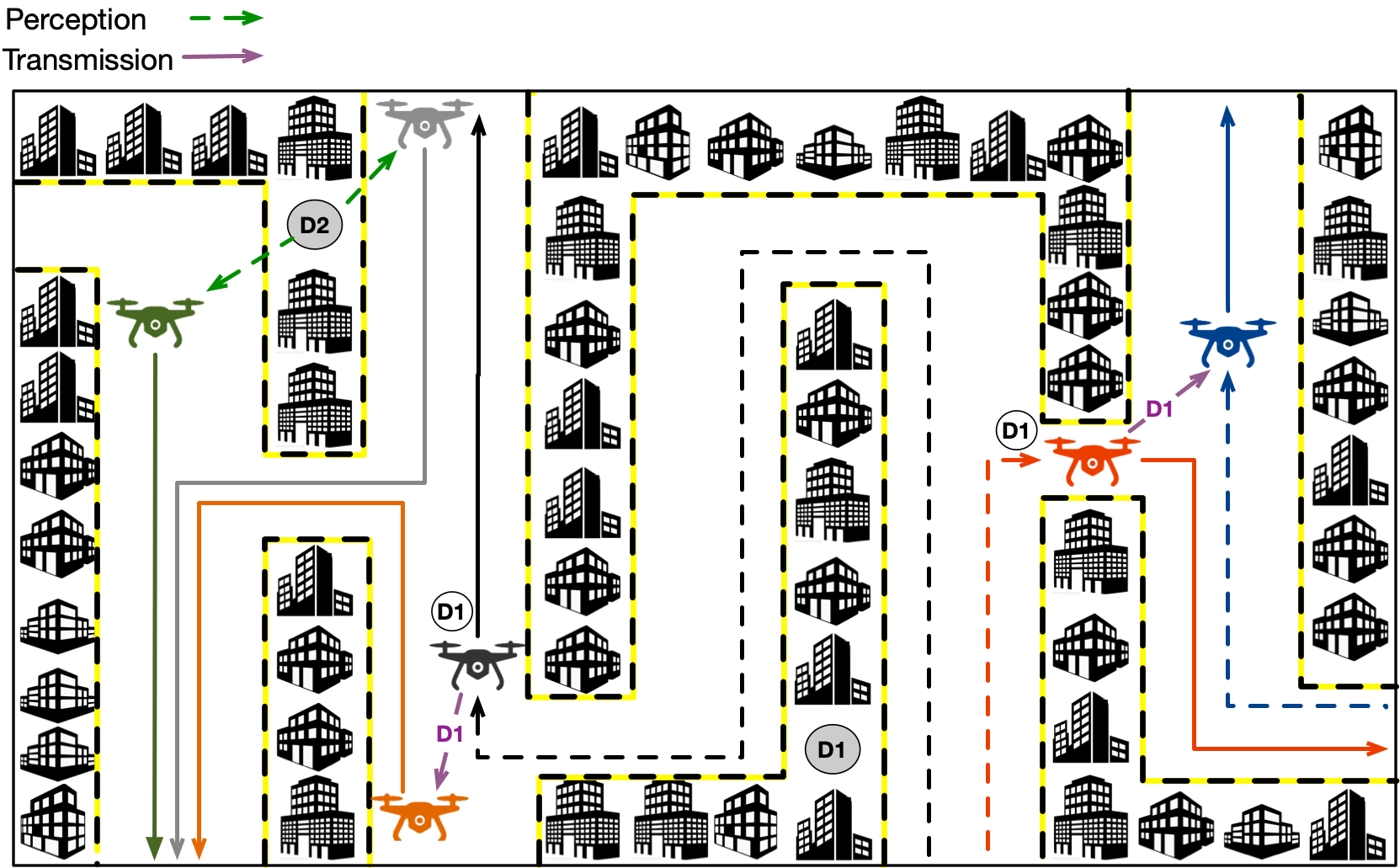}
    \caption{System Model. The grey circles represent the sources of the data referred to.}
    \label{fig:system_model}
    \vspace{-4mm}
\end{figure}

As shown in Figure \ref{fig:system_model}, our system consists of several UAVs $U$ deployed with communication capabilities that allow them to connect and transfer data among each other. Each of these UAVs is given a dedicated path in the area to flee to collect a set of information $I$ collaboratively. The paths are designed to reach all the information needed by the disseminated group of UAVs. Each of the UAVs must gather/receive a subset of $I$ (call it $I_u$) before leaving the area in order to send them to another UAV, or to analyze these data and make a decision (e.g., decide the upcoming path to take). All of these UAVs are assumed to be connected via a dedicated control channel to an agent (i.e., an edge server) responsible for establishing the data gathering and transfer plan. Each UAV has a range of communication. It cannot send data to another UAV unless it is in its range. For simplicity, this range is divided into several subranges. Each subrange is characterized by the transmission power required to transfer the data from the source UAV to a UAV located in this subrange. Consider the following transmission rate function:

\begin{equation}
    \label{shennonCapacity}
    \text{rate}(s, d) = \beta * \log_2(1 + \frac{\textit{Po} * \text{dist}(s, d)^{-\alpha}}{N_0 * \beta})\,,
\end{equation}

\noindent where $\beta$ is the bandwidth assigned to each channel, $\textit{Po}$ is the transmission power, $N_0$ is the thermal noise, $\alpha$ is the power loss decay factor and $\text{dist}(s, d)$ is the distance between source $s$ and destination $d$. We assume that all transmissions are assigned the same bandwidth, and the same white noise attenuates all. For simplicity, the timeline $T$ is divided into time units $\{t\}$. 
\vspace{-2mm}
\subsection{Connectivity Graph}
\begin{figure}[!htb]
    \vspace{2mm}
    \centering
    \fbox{\includegraphics[width = 0.85\columnwidth]{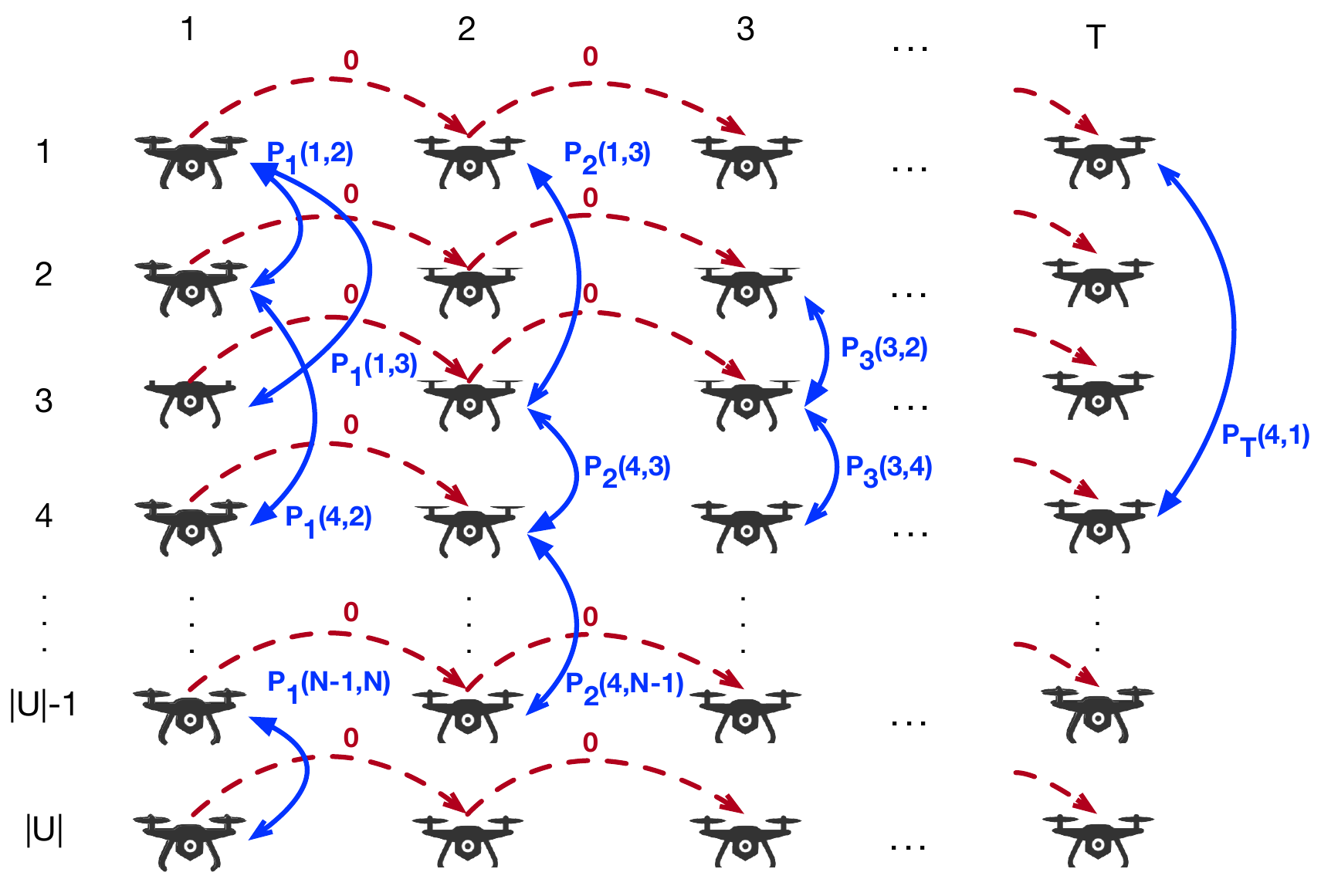}}
    \caption{UAVs connectivity graph.}
    \label{fig:UAV_Graph}
    \vspace{-6mm}
\end{figure}

We can structure the UAV's connectivity as a directed weighted graph. Let $\lambda_{uu'}^{tk}$ be a binary parameter that indicates that UAV $u'$ is in the $k^{\text{th}}$ subrange of UAV $u$ at time $t$. Let $r_u^k$ be the radius of $u$'s $k^{\text{th}}$ subrange. Let $\mathcal{G} = \{\vertices, \edges, \weights\}$ be the graph where $\vertices = \{(u,t) : u \in U \land t \leq T\}$,  $\edges = \{((u, t), (u',t')): (\exists k : \lambda_{uu'}^{tk} = 1 \land t = t') \lor (u = u' \land t' = t + 1)\}$ and $\weights = \{p_t(u, u'): ((u, t), (u',t)) \in \edges\}$, where $p_t(u, u')$ is the transmission power required to transmit a piece of information from $u$ to $u'$ at time $t$. A sample graph is shown in Figure \ref{fig:UAV_Graph}. We differentiate between two kinds of edges, the \textbf{\textit{connectivity edges}} (shown in blue) and the \textbf{\textit{caching edges}} (shown in red). The first is about having connectivity between two UAVs at a certain time unit. To pass through these edges, we need to pay a price, which is the transmission power required to perform the data transfer. Caching edges are a representation of the  UAV  caching data into the next time unit. Caching edges have no cost as they will not consume wireless resources. We will define the problem formally via the connectivity graph in the next subsection.

\subsection{Problem Definition}
\textbf{Problem $\mathsf{P}$ definition}: \textit{Given a connectivity graph $\mathcal{G} = \{\vertices, \edges, \weights\}$, where $\{\vertex \in \vertices\}$ is the set of its vertices, $\{ \edge \in \edges\}$ is the set of of edges and $\{\weight \in \weights\}$ is the set of edges weights. Let $\{i \in I\}$ be a set of information to collect. Some of the vertices, say $\vertices_i^s$, can generate information $i$ as their corresponding UAV can gather it. Let $\vertices_i^d$ be the set of information $i$'s destination. Each element of this set is a subset of $\vertices$, say $\mu(u)$, such that if one of the vertices in this subset received $i$, the entire subset is considered as it has received $i$ (the corresponding UAV has received $i$). An edge $\edge$ is called active if it is used to transfer a piece of information. Each edge can be used to send only one piece of information. An edges collision set is a subset of $\edges$ that only $C$ of them can be active the edges that exist in the same time unit). The goal is to activate a set of edges to transfer each piece of information from one of its sources to the destination at a minimum cost. The cost of the activation is the sum of maximum active vertices cost $\sum_{\vertex \in \vertices} \max_{\edge\in \edges_o^i(\vertex)} w_{\edge}$ where $\edges_o^i(\vertex))$ is the set of active outer edges of vertex $\vertex$.} 

We count only the edges with maximum weight in each vertex because increasing the transmission power to reach a certain point will allow the UAV to reach a closer point without consuming more power. 


\begin{theorem}
\label{prop:np_hard}
Problem $\mathsf{P}$ is a strong NP-hard problem.

\begin{proof}
Consider an instance $\Pi^{\text{steiner}}$ of the minimum Steiner tree problem (MST) \cite{proemelSteinerTree} where there is one tree source vertex and multiple destination vertices. The problem is to find the minimum partial spanning tree that reaches all the destinations. 
To reduce this problem to ours,
we create an instance of our problem, $\Pi^\text{uav}$, with only one collision set and only one information. 
The source in $\Pi^{\text{steiner}}$ should be a source of information, and the destinations should be the destinations of the information. 
In $\Pi^\text{uav}$, if we activate a higher weight edge, we also activate the lower ones without additional cost. To avoid that, after reduction from $\Pi^{\text{steiner}}$, for each vertex in it, we create a set of vertices with a size equal to the outer degree of the original vertex. 
Each vertex in the generated set should have the same inner edges as the original one but one outer edge from it. 
This will make any solution to $\Pi^\text{uav}$ a solution to $\Pi^{\text{steiner}}$. Since this reduction takes polynomial time and MST is NP-hard, we have our problem as NP-hard. 
\end{proof}
\end{theorem}

\section{Problem Formulation} \label{sec:ILP}
\begin{table}[!htb]
\vspace{-3mm}
\footnotesize
\caption{Symbols used in formulating the problem}
\centering
\begin{tabular}{|p {2.1cm} p{5.6cm}|}
\hline
\rowcolor{lightgray}
\textbf{Symbol}	& \textbf{Explanation}\\
\hline
\rowcolor{lightgray}
\multicolumn{2}{|l|}{\textbf{a) Parameters}}\\
\hline
$\vertex \in \vertices$ 						& Set of vertices\\
$\edge \in \edges$                              & Set of edges\\
$i \in I$ 						& Set of data\\
$\vertices_s^i$                                 & the source vertices of information $i$\\
$\vertices_d^i$                                   & the destination vertices of information $i$\\
$\edges_{o}(\vertex)$                            & the outer edges of vertex $\vertex$\\
$\edges_{i}(\vertex)$                            & the inner edges of vertex $\vertex$\\
$t \leq T$              & Time frames\\
$C$                     & The number of channels in the system.\\
\hline
\rowcolor{lightgray}
\multicolumn{2}{|l|}{\textbf{b) Variables}} \\
\hline
$P_{\vertex}$                         & the transmission cost of vertex $\vertex$\\
$\alpha_{\edge}^i$                    & edge $\edge$ is transmitting the piece of information $i$\\
$h_{\vertex}^i$                       & vertex $\vertex$ is a hub to transfer information $i$\\
$b_{\vertex}^i$                       & vertex $\vertex$ is transmitting information $i$\\
$d_{\vertex}^i$                       & vertex $\vertex$ is a destination of information $i$\\
\hline
\end{tabular}
\label{table:symbols}
\end{table}

 Table \ref{table:symbols} lists all the symbols used. The objective of our problem can be formulated as follows and it states that transmission cost should be minimized over all the vertices.
\begin{equation} \tag{OBJ}
    \label{ilp:obj}
    \min \sum_{\substack{\vertex \in \vertices}} P_{\vertex},. 
\end{equation}


\begin{enumerate}





\item If vertex $\vertex$ serves as a hub to transfer information $i$, then at least one outer edge should be active transferring $i$. If the hub variable is active, then C2 says it has at least one outer edge, and C1 limits it to have the maximum possible outer edges. 
\begin{constraint} 
\begin{aligned}
    \label{ilp:c2}
    \sum_{\edge \in \edges_o(\vertex)} \alpha_{\edge}^i  \leq |\edges_{o}(\vertex)|  h_{\vertex}^i\ \ \forall \vertex \in \vertices/\vertices_s^i \ \ \forall i \in I\,.
\end{aligned}
\end{constraint}
\vspace{-2mm}
\begin{constraint} 
\begin{aligned}
    \label{ilp:c21}
    \sum_{\edge \in \edges_o(\vertex)} \alpha_{\edge}^i  \geq  h_{\vertex}^i\ \  
    \forall \vertex \in \vertices/\vertices_s^i \ \ \forall i \in I\,.
\end{aligned}
\end{constraint}

\item If vertex $\vertex$ is a hub to transfer information $i$, then C3 says, one of its inner edges should be active, transferring $i$. While C4 incorporates the case where a destination can be a hub too in a different time unit, hence it should have at least one active incoming edge.
\vspace{-1mm}
\begin{constraint} 
\begin{aligned}
    \label{ilp:c2}
    \sum_{\edge \in \edges_i(\vertex)} \alpha_{\edge}^i  = h_{\vertex}^i\ \ \forall \vertex \in \vertices/\vertices_d^i\ \forall i \in I\,.
\end{aligned}
\end{constraint}
\vspace{-2mm}
\begin{constraint} 
\begin{aligned}
    \label{ilp:c2}
    \sum_{\edge \in \edges_i(\vertex)} \alpha_{\edge}^i  = h_{\vertex}^i  \lor d_{\vertex}^i\ \ \forall \vertex \in \vertices_d^i \ \forall i \in I\,.
\end{aligned}
\end{constraint}


\item For each subset $\mu(u)$, for each required information $i$, one of $\mu(u)$ should receive $i$. The destination variable activates once to receive i in a specific time unit.
 \vspace{-1mm}
\begin{constraint} 
\begin{aligned}
    \label{ilp:c3}
    \sum_{\vertex \in \mu (u)} d_{\vertex}^i  = 1\ \ 
    \forall i \in I\ \forall u \in U_d^i\,.
\end{aligned}
\end{constraint}

 

\item One of information $i$ sources, at least, should send the information. 
\vspace{-1mm}
\begin{constraint} 
\begin{aligned}
    \label{ilp:c3'}
    \sum_{u \in \vertices_s^i} \sum_{\edge \in \edges_o(\vertex)} \alpha_{\edge}^i  \geq 1\ \ 
    \forall i \in I\,.
\end{aligned}
\end{constraint}


\item A vertex $\vertex$ can only transmit one information. C7 restricts a node from transferring more than one piece of information at a time. While C8 restricts the maximum outer edges.
\vspace{-0.25mm}
\begin{constraint} 
\begin{aligned}
    \label{ilp:c4}
    \sum_{i \in I} b_{\vertex}^i \leq 1\ \ 
    \forall \vertex \in \vertices\,.
\end{aligned}
\end{constraint}
\vspace{-3mm}
\begin{constraint} 
\begin{aligned}
    \label{ilp:c41}
    \sum_{\edge \in \edges_o(\vertex)} \alpha_{\edge}^i \leq 
    |\edges_{o}(\vertex)| \times 
    b_{\vertex}^i\ \ 
    \forall \vertex \in \vertices\ \forall i \in I\,.
\end{aligned}
\end{constraint}

\item The number of active edges in a collision set can not exceed C. 
\vspace{-1mm}
\begin{constraint} 
\begin{aligned}
    \label{ilp:c5}
    \sum_{\edge \in \edges_t} \sum_{i \in I} \alpha_{\edge}^i \leq C\ \ 
    \forall t \leq T\,.
\end{aligned}
\end{constraint}


\item The transmission cost of a vertex $\vertex$ is the maximum weight of its active outgoing edges, accounting for energy consumption by other active edges.
\vspace{-1mm}
\begin{constraint} 
\begin{aligned}
    \label{ilp:c6}
    P_{\vertex} \geq \weight \times \sum_{i \in I} \alpha_{\edge}^i\ \ 
    \forall \edge \in \edges_o(\vertex)\ \ \forall \vertex \in \vertices\,.
\end{aligned}
\end{constraint}
\end{enumerate}

\section{Heuristic Algorithms} \label{sec:heuristics}
\begin{figure}[!htb]
    \centering
    \fbox{\includegraphics[width=0.9\linewidth]{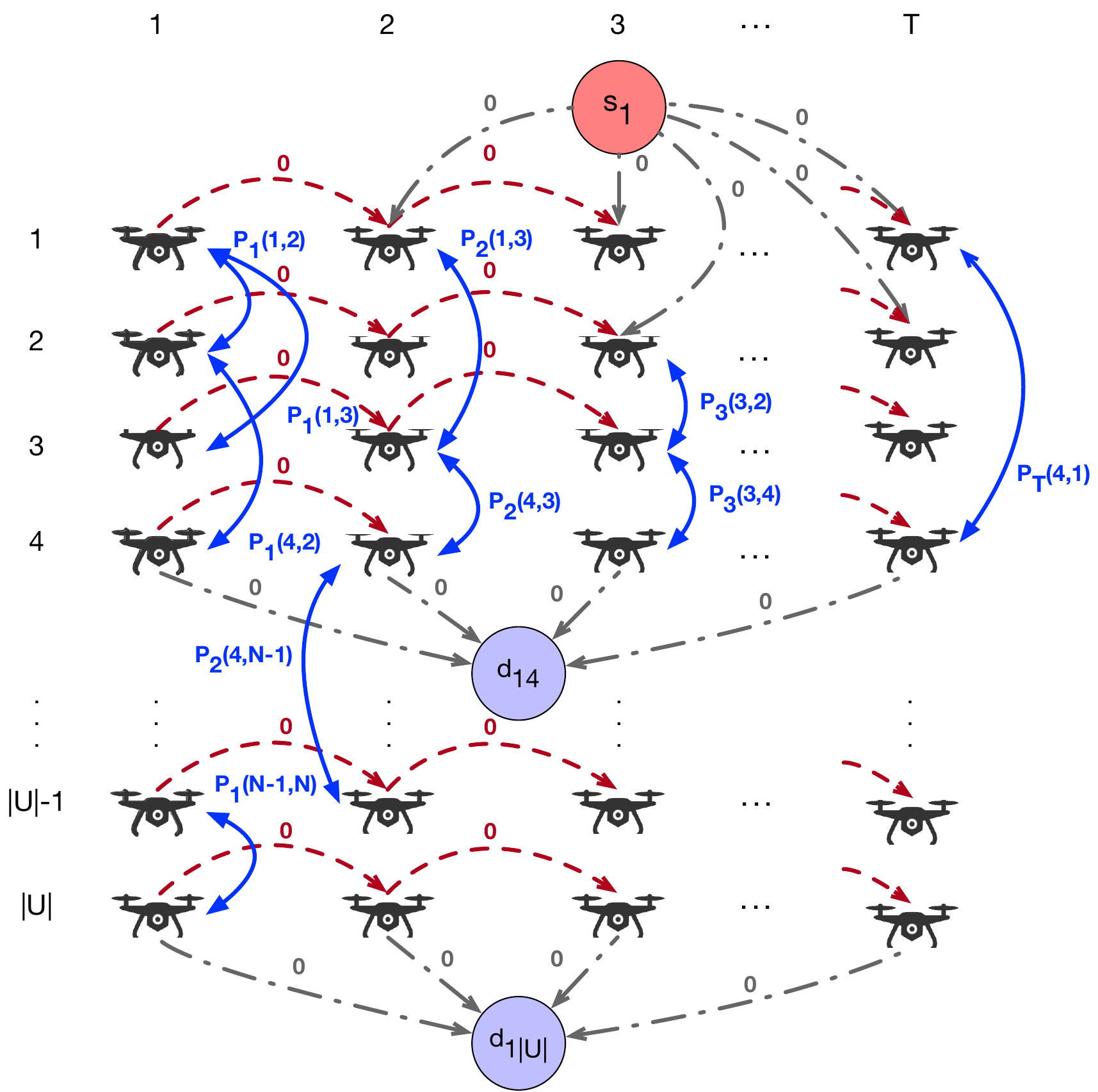}}
    \caption{The augmented graph of UAVs. $s_1$ is the virtual source of information $i$ and $d_{14}$, $d_{1|U|}$ are the virtual destinations of information $i$ that corresponds to UAV 4 and $|U|$, respectively.}
    \label{fig:augmented}
    \vspace{-0.5mm}
\end{figure}

This section proposes greedy algorithms that leverage heuristics to sort information. As illustrated in Figure \ref{fig:UAV_Graph}, we modeled the problem as a graph, as explained in Section \ref{sec:systemModel}. To implement graph theory-based solutions, we also adopted the concept of virtual vertices. We connected all sources of a specific piece of information (across the timeline) to a single virtual vertex and assigned all new edges a zero weight. Similarly, we linked all vertices corresponding to the same UAV over time to another virtual vertex (see Figure \ref{fig:augmented}). This approach allows us to focus on one source vertex for each piece of information and one destination vertex for each destination UAV. By applying this method, we can utilize off-the-shelf graph theory solutions to address the problem. From our initial insights, it is clear that each information dissemination forms a Steiner tree among the source and the destinations. The minimum Steiner tree problem is well-known as NP-hard, and attempting to implement an exact solution would result in exponential execution time. Therefore, we implemented a one-source-to-multiple-destination shortest path solution using Dijkstra's algorithm \citep{proemelSteinerTree}
 to approximate the trees. The details of this solution are in Algorithm 1.

\begin{figure}[!htb]
    \centering
    \includegraphics[width=\linewidth]{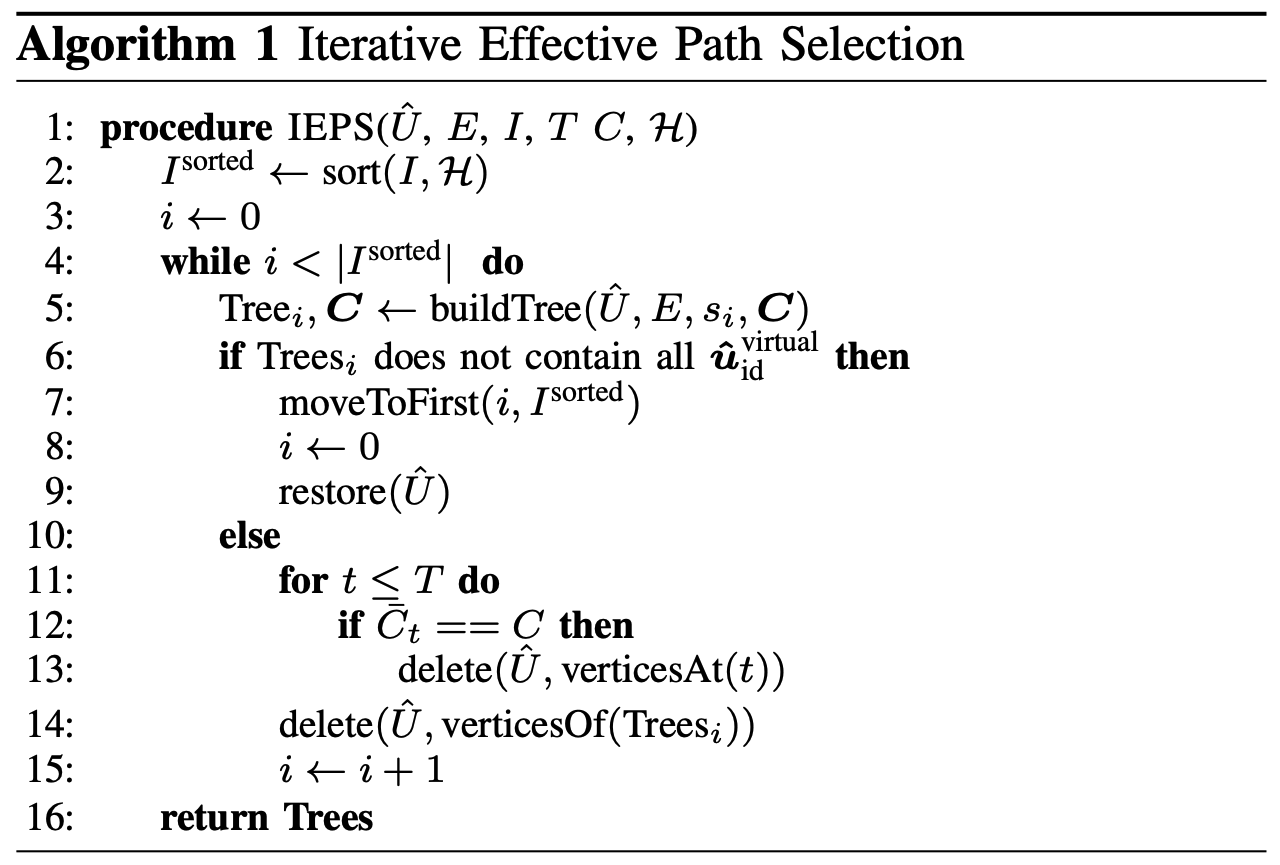}
\end{figure}

\begin{figure}[!htb]
    \vspace{3mm}
    \centering
    \includegraphics[width=\linewidth]{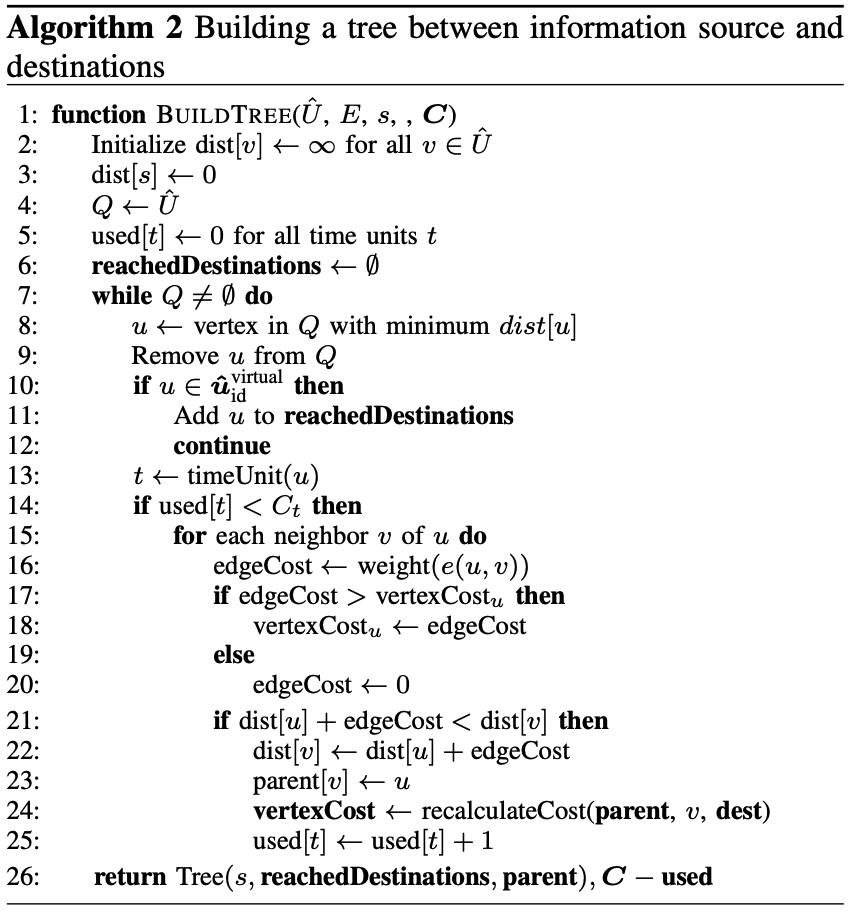}
    \vspace{-6mm}
\end{figure}

The first input is $\hat{U}$, which is $\tilde{U}$ augmented with virtual vertices. The algorithm starts by sorting the information based on the heuristic provided (line 1). We considered four heuristics for sorting the information: Most Power First (MPF), Least Power First (LPF), Most Number of UAVs First (MUF), and Random (R). Lines 2-15 include the primary iteration. It goes through the sorted information (line 3) and serves them individually. For each, it calculates the tree from the virtual source to all the virtual destinations using the \textit{BuildTree} function (line 5). If a tree is found (line 6), it deletes all the used vertices to avoid conflict with the not-yet-served information (line 14). If the number of channels in a certain time unit has reached its limit, all the vertices in this time unit get deleted, so no information is considered in this time unit to pass any info (lines 11-14). 
If an information tree couldn’t be established, the algorithm restarts the primary iteration by restarting the counter. Also, it pushes the information found challenging to serve to the beginning of the list to improve its chance of finding enough resources (lines 7-8).
Algorithm 2 finds the shortest paths between the information's source and destinations while considering the number of channels. It calculates the cost by counting only the maximum vertices' active outer edges.
\vspace{-2mm}
\section{Performance Evaluation} \label{sec:perEval}
To assess our approaches, we conducted performance evaluations on simulated scenarios. Our method
were tested using procedurally generated problem instances. We developed the algorithms in C++ and the ILP using CPLEX (an optimization tool by IBM). 
We used the following values: $\beta$: 40MHz, $\alpha$: 2, $N_0$: $1e-9$, packet size: 200KB, $T$: 200, ranges: 10 and time units lengths: 0.01s.

\begin{figure} [!htb]
    \centering
    \fbox{\includegraphics[width=0.9\linewidth]{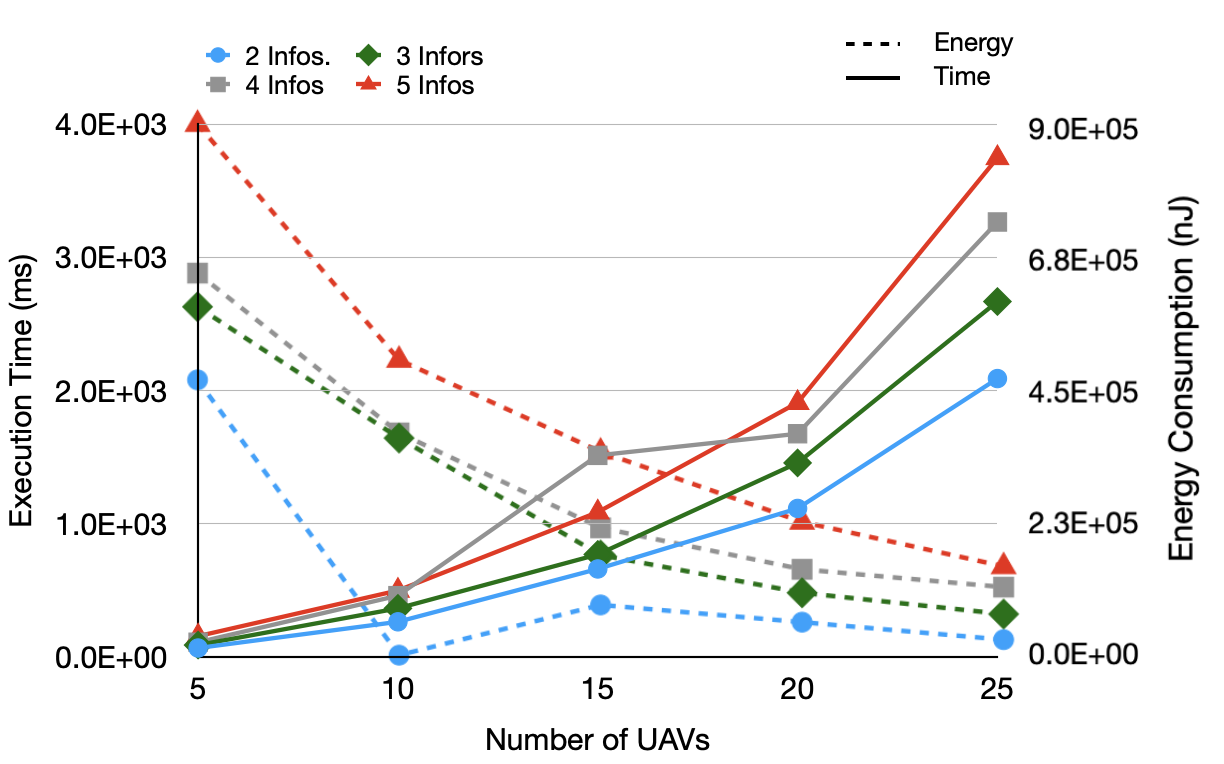}}
    \caption{Algorithm performance}
    \label{fig:algoPerf}
\end{figure}

Table \ref{table:performance} compares the performance of the heuristics and CPLEX. The limited instance size is due to the incapability of CPLEX to handle larger instances. All heuristics are faster than CPLEX, with a minimum speed-up of 99\%. MPF performs best in terms of the objective. 
The reason is when we remove vertices already used to serve information, data requiring high power consumption must be rerouted to reach all destinations. This rerouting increases resource consumption and power usage. By prioritizing information with higher power consumption, MPF reduces the likelihood of increased power consumption due to rerouting. Figure \ref{fig:algoPerf} shows that more infos generally will cause more power consumption. Additionally, computation time increases with UAVs and infos.

\begin{table*}[!htb]
\vspace{10mm}
\centering
\caption{Performance Comparison. ET: execution time (ms). Dev.: Deviation from optimal. Obj: energy consumption (nJ).}
\label{table:performance}
\resizebox{0.9\textwidth}{!}{%
\begin{tabular}{|>{\columncolor[gray]{0.9}}c|>{\columncolor[gray]{0.9}}c|>{\columncolor[gray]{0.9}}c|cc|ccc|ccc|ccc|ccc|}
\hline
\rowcolor[gray]{0.9}
\textbf{$|U|$} & \textbf{$|I|$} & \textbf{$T$} & \multicolumn{2}{c|}{\textbf{ILP}} & \multicolumn{3}{c|}{\textbf{R}} & \multicolumn{3}{c|}{\textbf{MPF}} & \multicolumn{3}{c|}{\textbf{LPF}} & \multicolumn{3}{c|}{\textbf{MUF}} \\
\cline{4-17}
\rowcolor[gray]{0.9}
&  &  & \textbf{Obj} & \textbf{ET} & \textbf{Obj} & \textbf{Dev.} & \textbf{ET} & \textbf{Obj} & \textbf{Dev.} & \textbf{ET} & \textbf{Obj} & \textbf{Dev.} & \textbf{ET} & \textbf{Obj} & \textbf{Dev.} & \textbf{ET} \\
\hline
4 & 2 & 40 & 13150.00 & 8549.16 & 16950.17 & 46.42\% & 2.19 & 15675.00 & 25.61\% & 2.03 & 17225.17 & 52.98\% & 2.01 & 16950.17 & 46.42\% & 2.01\\
4 & 2 & 60 & 16137.50 & 65132.79 & 20725.28 & 39.92\% & 4.22 & 19287.50 & 28.35\% & 3.82 & 20862.80 & 44.26\% & 3.89 & 20725.28 & 39.92\% & 3.82\\
5 & 2 & 40 & 12026.79 & 94716.78 & 14424.46 & 20.38\% & 3.06 & 13982.14 & 14.91\% & 2.96 & 14585.20 & 21.05\% & 3.08 & 14424.46 & 20.38\% & 2.98\\
5 & 2 & 60 & 11417.76 & 266741.63 & 15256.92 & 40.09\% & 5.83 & 14230.26 & 28.43\% & 5.53 & 15424.68 & 42.02\% & 5.57 & 15256.92 & 40.09\% & 5.93\\
6 & 2 & 40 & 11017.24 & 186986.85 & 14457.38 & 31.75\% & 8.06 & 14534.48 & 29.67\% & 8.11 & 13888.45 & 30.29\% & 6.45 & 14457.38 & 31.75\% & 8.01\\
6 & 2 & 60 & 9708.33 & 347091.27 & 12125.50 & 29.59\% & 8.06 & 12111.11 & 28.69\% & 8.18 & 12389.43 & 32.4\% & 7.64 & 12125.50 & 29.59\% & 7.93\\
\hline
\multicolumn{3}{|c|}{Average} & 12242.94 & 161536.41 & 15656.62 & 34.69\% & 5.24 & 14970.08 & 25.94\% & 5.10 & 15729.29 & 37.17\% & 4.77 & 15656.62 & 34.69\% & 5.11\\
\hline
\end{tabular}%
}
\vspace{-5mm}
\end{table*}

\begin{figure} [!htb]
    \centering
    \fbox{\includegraphics[width=0.9\linewidth]{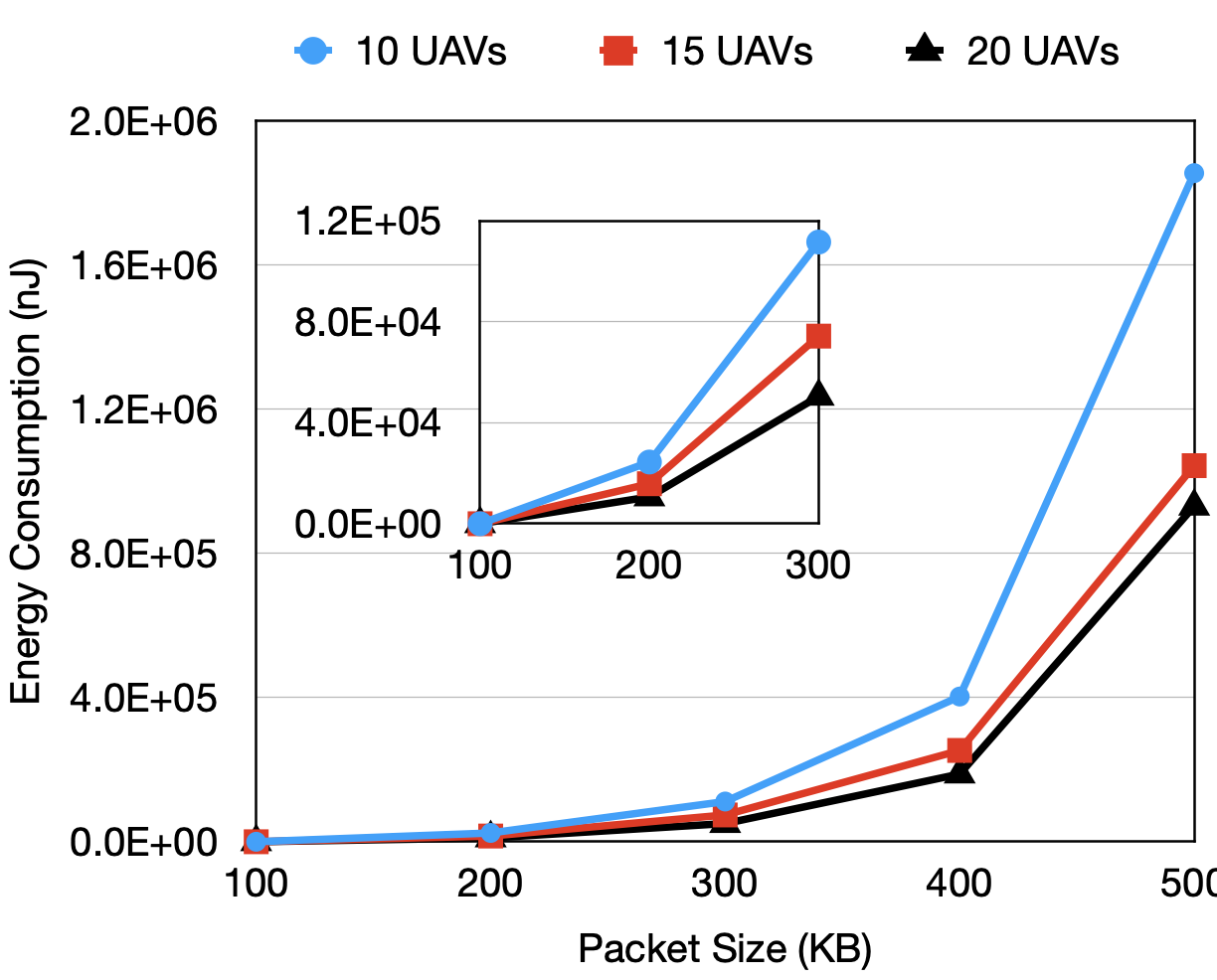}}
    \caption{The energy consumption level vs the packet size.}
    \label{fig:PacketSize}
\end{figure}

\begin{figure}[!htb]
    \centering
    \fbox{\includegraphics[width=0.9\linewidth]{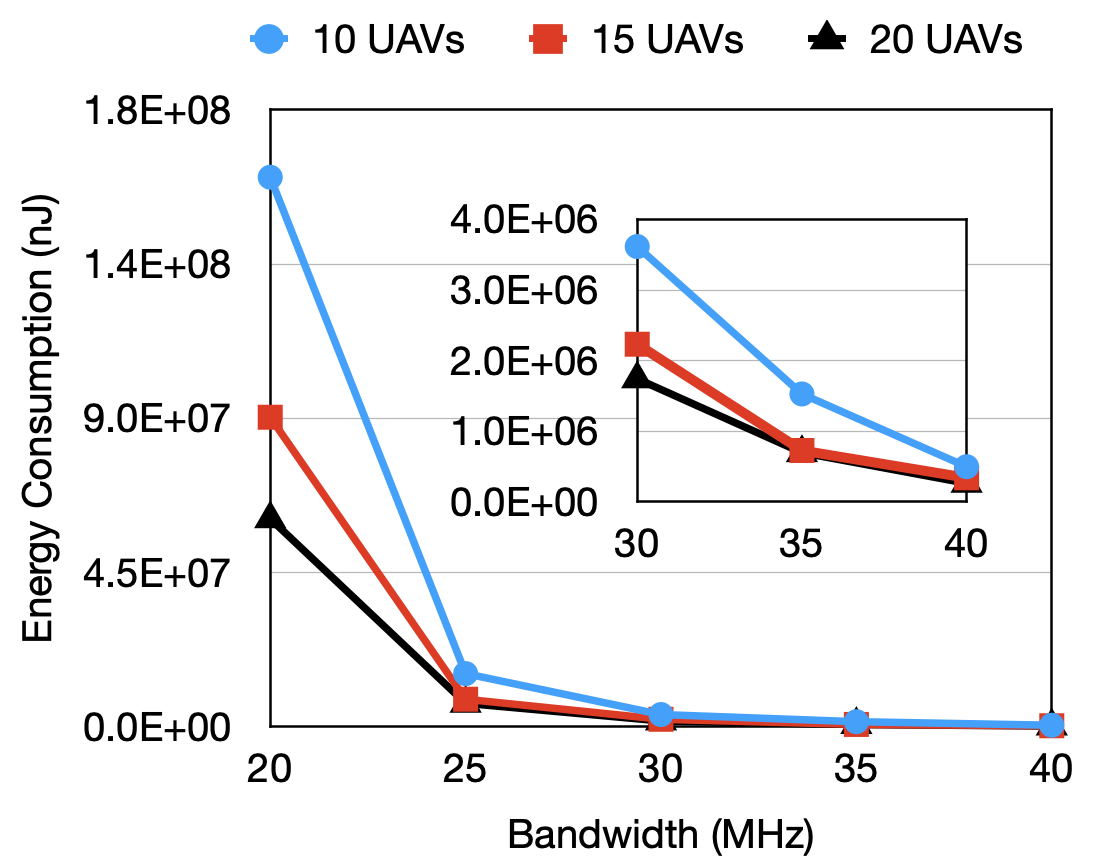}}
    \caption{The energy consumption level vs the bandwidth.}
    \label{fig:bandwidth}
    \vspace{-5mm}
\end{figure}

Figures \ref{fig:PacketSize} and \ref{fig:bandwidth} highlight some UAV system energy dynamics using MPF. The consumption increases exponentially with packet size.
Conversely, higher bandwidth reduces energy usage as expected. Notably, larger UAV networks consistently demonstrate improved energy efficiency across both variables. This trend suggests that increasing the number of UAVs leads to better load distribution. This underscores the importance of optimizing packet size, bandwidth, and network size in collaborative UAV-based data-gathering systems to achieve high energy efficiency.

\vspace{-6mm}


\section{Conclusion}
We presented an approach to collaborative UAV-based data gathering and sharing in IoT environments, addressing the challenges of optimizing communication overhead. Our contributions include a system model accounting for UAV mobility and multi-hub communication paths, an ILP formulation, heuristic algorithms for efficient routing, and experimental evaluation. The results demonstrate the effectiveness of collaborative UAV networks in reducing power consumption while performing all requested data transfers. By addressing the challenge of limited power, our approach paves the way for more efficient and reliable UAV-based data-gathering systems, representing a significant step forward in harnessing the potential of UAVs for IoT data collection. In our future work, we will consider overcoming node/link failures, which is essential to improving the robustness of a dynamic wireless communication network. 
\vspace{-1mm}

\bibliographystyle{ieeetr}
\bibliography{references}

\end{document}